%% file: proc.tex
\documentclass[fleqn,twoside]{article}
\usepackage{my_espcrc2,isolatin1}

\usepackage{epsf}
\usepackage[figuresright]{rotating}

\newcommand{\AmS}{{\protect\the\textfont2
  A\kern-.1667em\lower.5ex\hbox{M}\kern-.125emS}}

\usepackage{fancybox,cite}
\include{macros}

\include{titlehep}
       
\begin{document}

\begin{abstract}
\noindent
After a short presentation of lattice QCD and some of its current
practical limitations, I review recent progress in applications to
phenomenology. Emphasis is placed on heavy-quark masses and on
hadronic weak matrix elements relevant for constraining the CKM
unitarity triangle. The main numerical results are highlighted in
boxes.
\vspace{-0.3cm}
\end{abstract}

\maketitle

\section{Introduction}

The lattice formulation of quantum field theory, combined with large
scale numerical simulations, is contributing in a variety of ways to
the current research effort in particle physics. In the present talk I
will focus on lattice QCD and its rôle in quantifying non-perturbative
strong interaction effects. I wish to apologize straight away to
those colleagues whose work I will not be able to cover.

The objectives of lattice QCD are numerous, beginning with the goal of
establishing QCD as the theory of the strong interaction in the
non-perturbative domain. This can be
achieved by comparing quantities calculated on the lattice, such as
hadronic spectra, deep inelastic scattering structure functions,
etc.\ to their experimentally measured counterparts.  Some of these
aspects were reviewed in the parallel session talk by Collins
\cite{sara} and I will not repeat them here. Lattice QCD can also be used
to determine the fundamental parameters of QCD, such as the strong
coupling $\alpha_s$ or the quark masses. Because significant progress
has been made in the evaluation of heavy-quark masses in the last
year, I will devote a section to them. I refer you to the talk
of Wittig at Lattice 2002 for a discussion of light-quark masses
\cite{Wittig:2002ux}.

Another important objective of lattice QCD is to evaluate the
non-perturbative, strong-interaction corrections to weak processes
involving quarks. Indeed, our inability to reliably quantify these
long-distance effects is often the dominant source of uncertainty in
measuring fundamental weak-interaction parameters, such as the
elements of the Cabibbo-Kobayashi-Maskawa (CKM) matrix. This problem
has become all the more acute that the results from the CESR, LEP, the
Tevatron and now the $B$ factories indicate that deviations from the
standard model, if present, must be small. Since this is a natural
meeting point between lattice QCD and a significant fraction of the
audience, I will devote the largest part of my talk to the review of
calculations of weak matrix elements relevant for constraining the
unitarity triangle. This area is one where the interplay of lattice
QCD and the rich experimental program of the next few years will be
strong, with lattice QCD in turn contributing and improving thanks to
the many precise measurements to come. In this exchange, one should
not overlook the important rôle that CLEO-c will play, by providing
the lattice community with results of unprecedented accuracy, for
$D$-meson decays in particular. Not only will these results provide
stringent tests of the lattice method in the heavy-quark sector, but
they will allow us to normalize our $b$-physics results by the
equivalent charmed quantities, thereby reducing many of our systematic
errors. I will unfortunately not have the time to discuss $D$-meson
processes here and I refer you to the reviews of
\cite{Bernard:2000ki,Ryan:2001ej} for recent summaries.

The discussion up to now certainly does not exhaust the range of
subjects studied in lattice QCD. One subject of great phenomenological
interest, which is not covered here, is that of non-leptonic kaon
decays and direct $CP$ violation in these decays. While significant
progress, both theoretical and numerical, has been made in the last
years (see e.g. the reviews
\cite{Ishizuka:2002nm,Sachrajda:2001ey,Lellouch:2000bm}), the results
have not yet reached a level of maturity which permits detailed comparison
with experiment. Another important subject is the rôle that lattice QCD
plays in understanding QCD and finite temperature and/or density. This
aspect was covered by Alford in his plenary lecture at this
conference \cite{mark}. Lattice QCD can also be used to make
predictions for exotic hadrons, such as $b\bar b g$ (see
e.g. \cite{sara}), and provides a powerful tool for understanding the
mechanism(s) of confinement and spontaneous chiral symmetry breaking.

The remainder of the talk is organized as follows. In
\sec{sec:challenges}, I briefly review the lattice method and some of its
current practical limitations. In \sec{sec:hqmasses}, I discuss new
developments in the computation of the charm and bottom-quark
masses. I then review recent progress in the calculation of matrix
elements relevant for constraining the CKM unitarity triangle in
\sec{sec:ckmology}~\footnote{Please see also the parallel session talks
of D.~Becirevic \cite{damir} and J.~Simone \cite{jim}.} and provide an
outlook in \sec{sec:ccl}.

\section{The challenges of lattice calculations}

\label{sec:challenges}

\begin{figure}
\centerline{\epsfxsize=5.cm\epsffile{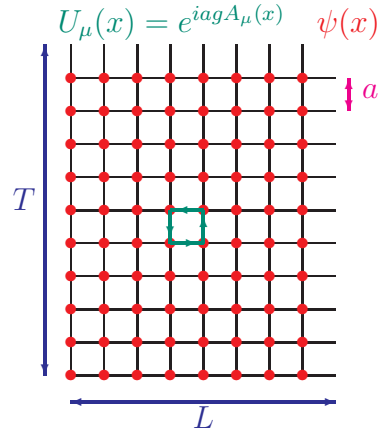}}
\vspace{-1.cm}
\caption{\em Two dimensional projection of a four dimensional cubic lattice.
$a$ is the lattice spacing.}
\vspace{-0.6cm}
\label{fig:latfig}
\end{figure}

Before reviewing some of the limitations of present day zero-temperature
and zero-density lattice
calculations, let me first say a few words about why lattice QCD is
needed at all. As all of you know, quarks and gluons are confined
within hadrons.  Since confinement
cannot be explained in perturbation theory, a non-perturbative tool is
required to relate experiment to the underlying theory. Lattice gauge
theory is a good candidate, as it provides a mathematically sound
definition of non-perturbative QCD. In lattice QCD, euclidean
spacetime is approximated by a discrete set of points. Quark fields
reside on these points and gauge fields on the links between the
points, as shown in
\fig{fig:latfig}. When the lattice is finite, 
this discretization provides both
ultraviolet and infrared cutoffs and yields a well defined path
integral.  Because the number of degrees of freedom is finite and the
integrand positive, this
integral can be evaluated numerically using stochastic
methods. Lattice QCD thus allows the computation of hadronic
observables directly from the QCD lagrangian, i.e. from {\it first
principles.} The only source of errors are the finite statistics and
the discretization of spacetime. The former decrease with the
square-root of the number of samples generated. The latter are
proportional to powers of $a\lqcd$ and $ap_\mu$, where $a$ is the
lattice spacing (see
\fig{fig:latfig}), $\lqcd$ is the QCD scale and $p_\mu$ denotes 
the four-momenta of
the particles studied. The two errors can be made arbitrarily small,
by increasing the statistics and by reducing the lattice spacing. 

This is how lattice QCD works in principle. Let me now review
some of the main difficulties that we encounter in practice.

\subsection{Numerical cost and quenching}

Though the number of degrees of freedom is finite, it is still very
large. On a $32^3\times 64$ lattice, necessary to have a box of
three-and-some fermi with an inverse lattice spacing of $2\,\gev$, the
number of degrees of freedom is of order $10^8$. In addition, the cost
of including sea-quark effects is very high and increases rapidly as
the sea-quark mass is reduced. This explains the ``popularity'' of the
quenched approximation, where valence quarks are treated exactly and
sea-quark effects are approximated by a mean field. I will denote this
approximation by $N_f=0$, where $N_f$ is the number of sea quarks
included in the calculation. The benefit of this approximation, or
rather truncation, is a large reduction in numerical cost since it
saves calculating the very numerically intensive fermion
determinant. The down side is that there is no small parameter which
controls the approximation so that the size of corrections is not
known {\em a priori}. Experience indicates that in many circumstances
the error induced does not exceed 10 to 15\%, as is shown, for
instance, for the case of light-hadron masses
\cite{Aoki:2002fd}. It is clear, however, that for processes which
depend dominantly on the existence of sea-quark effects, the quenched
approximation will be very poor.

While significantly more costly, more and more unquenched or partially
quenched~\footnote{Partially quenched refers to calculations where
valence and sea quarks are allowed to be different. Real world
QCD is thus a special case of partial quenching in which these quarks
are identified.}  calculations are being performed. Most of them are
done with two flavors of sea quarks ($N_f=2$), whose masses are
usually substantially larger than those of the physical up and down
quarks, instead of the three ($u$, $d$ and $s$, $N_f=3$) of our
world. Furthermore, because we have much less experience with these
calculations, we are still learning how to control their systematic
errors.

\subsection{Light quarks and chiral extrapolations}

Present day algorithms become much less effective when the mass $m_q$
of light quarks is reduced. At the same time, unwanted finite volume
effects increase. Thus at present, $N_f=2$ calculations are usually
limited to $m_q\gsim m_s/2$ or $M_\pi^{lat}\gsim M_K^{expt}$, where
$m_s$ is the strange-quark mass. This is a regime where the ``$\rho$''
cannot decay into two ``pions''. One notable exception is the $N_f=3$
simulation by MILC where quark masses of about $0.2m_s$ are reached
(see
\cite{Bernard:2001wy,Bernard:2002ep}, for instance). In any case, to
reproduce the physics of $u$ and $d$ quarks, observables computed for
values of $m_q>m_{u,d}$ must be {\it chirally extrapolated} to
$m_q=m_{u,d}$. Can this extrapolation be performed in a controlled
fashion? The only model-independent theoretical guide that we have is
chiral perturbation theory ($\chi$PT). The question then becomes
whether the quark masses used in simulations are light enough for
$\chi$PT at low order to be applicable. The question of chiral
extrapolations will be discussed more below. It was also the subject
of a panel discussion at Lattice 2002 \cite{Bernard:2002yk}.

\subsection{Heavy quarks and lattice artefacts}

\label{sec:hqanddisc}

The discretization errors associated with a heavy quark are
proportional to powers of $am_Q$, where $m_Q$ is the mass of this
quark. For these errors to remain under control, $am_Q$ must be much
smaller than 1, or $m_Q\ll a^{-1}$. Since present day inverse lattice
spacings are in the range of 2 to 4 GeV, it is clear that
the $b$ quark, with $m_b\simeq 5\,\gev$, cannot be simulated directly.
A number of solutions to this problem exist:

\smallskip

$\bullet$ {\it Relativistic quarks:} calculations are performed for
a number of heavy-quark masses around that of the charm, where
discretization errors are under reasonable control, and the results
are extrapolated in powers of $1/m_Q$ up to the $b$-quark mass, using
heavy quark effective theory (HQET) as a guide. The drawback of this
approach is that the extrapolation can be significant and that
discretization errors proportional to powers of $am_Q$ may be
amplified if this extrapolation is performed before a continuum
limit is taken.

\smallskip

$\bullet$ {\it Effective theories:} for heavy quarks whose mass $m_Q$
is large compared to the other scales in the problem, denoted here by
$\mu_{QCD}$, an expansion of QCD in powers of $\mu_{QCD}/m_Q$ can be
performed. There are different implementations of this idea which go
by the names of lattice HQET, of which the leading term is the static
approximation, NRQCD and Fermilab approach. The important benefit of
these approaches is that discretization errors are proportional to
powers of $a\mu_{QCD}$ instead of powers of $am_Q$. The drawback is
that precise calculations at the physical $b$-quark mass require the
calculation of corrections proportional to powers of $1/m_b$, which
are difficult to renormalize accurately
\cite{Bernard:2000ki}. In addition, the number of operators whose
matrix elements must be computed grows rapidly with the power of
$1/m_b$.  In the case of NRQCD, one is also confronted with the fact
that the continuum limit cannot be taken.

\smallskip

$\bullet$ {\it Combination of relativistic and HQET results}: in this
case, the extrapolations from the charm and infinite heavy-quark-mass
regimes are replaced by an interpolation. This is clearly a very good
way to reach the $b$ quark. Discretization and renormalization errors
are, however, very different in the two theories.  These
interpolations are thus really only reliable if the results in the two
theories are obtained in the continuum limit and are
non-perturbatively renormalized.

\section{Heavy-quark masses}

\label{sec:hqmasses}

Significant progress has been made in the determination of the charm
and $b$-quark masses in the last year. Unfortunately, the numerical
work is still performed in the quenched approximation. 

\subsection{The charm-quark mass}

The charm-quark mass is determined by tuning the bare heavy-quark mass
in the calculation until the experimental value of an observable, such
as the mass of the $D_s$ meson or the spin-averaged mass of the $1S$
$c\bar c$ states, is reproduced. The bare mass is then matched onto a
continuum renormalization scheme.

The results of two new calculations appeared this year. The first
\cite{Becirevic:2001yh} is a modern version of the earlier work of
\cite{Gimenez:1998uv}. While the latter made use of Wilson fermions,
the new calculation is performed with $\ord{a}$-improved Wilson
fermions, for which discretization errors are reduced from
$\ord{am_c}$ to $\ord{(am_c)^2}$. This is important here since
$am_c\sim 0.4$. Another improvement is the use of
next-to-next-to-next-to-leading-log ($N^3LL$) results to convert the
mass obtained non-perturbatively in the RI/MOM scheme at $\sim
3\,\gev$ to its renormalization-group invariant (RGI)
\cite{Chetyrkin:1999pq} and $\msbar$ values 
\cite{Chetyrkin:1997dh,Vermaseren:1997fq}. This conversion
had previously only been performed at $NLL$ order.  The calculation is
performed in the quenched approximation at a single value of
$a^{-1}\sim 2.7\,\gev$. The input used to fix the charm-quark mass is
the $D_s$ meson mass, with $m_s$ determined from $M_K$
and $a^{-1}$ from $M_{K^*}$. A number of systematics are considered
and the result obtained is: $\mmsbar_c(\mmsbar_c)=1.26(3)(12)\,\gev$.

\begin{figure}[t]
\centerline{\epsfxsize=7.5cm\epsffile{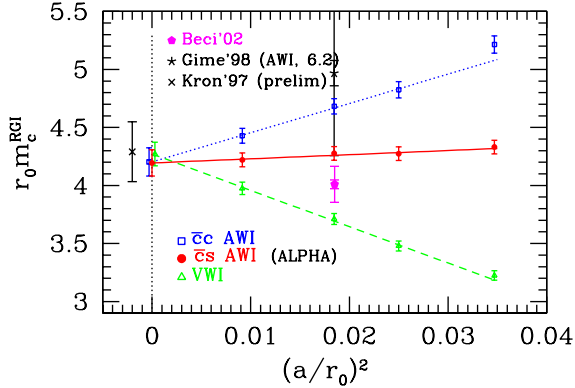}}
\vspace{-1.0cm}
\caption{\em Continuum extrapolation of the RGI charm mass 
by ALPHA 
\cite{Rolf:2002gu}. All
quantities are plotted in units of the scale $r_0\simeq 0.5\,\fm$
\cite{Sommer:1993ce}. The vertical dashed line corresponds to the
continuum limit. The other results
are: Beci'02
\cite{Becirevic:2001yh}, Gime'98 \cite{Gimenez:1998uv} and Kron'97 
\cite{Kronfeld:1997zc}. 
Not shown is a result obtained from combining sumrule techniques with
lattice results
\cite{Bochkarev:1995ai}.}
\vspace{-0.5cm}
\label{fig:mc}
\end{figure}

The second calculation \cite{Rolf:2002gu} is presumably the ultimate
quenched $m_c$ calculation. It is also the first time the continuum
limit has been thoroughly studied for this quantity.  The calculation
is performed at four values of the lattice spacing, ranging from
approximatively 2 to 4 GeV and makes use of $\ord{a}$-improved
fermions. The input used is similar to the calculation of
\cite{Becirevic:2001yh}. Five different definitions of the bare quark
mass are considered. They are based on the vector Ward identity (VWI),
the $\bar cs$-axial Ward identity (AWI) and a non-singlet $\bar
cc$-AWI. The renormalized masses obtained from these definitions
differ only by discretization errors and should therefore agree in the
continuum limit. The renormalization is performed non-perturbatively à
la ALPHA
\cite{Capitani:1998mq}, which means that the RGI mass 
is obtained without relying on perturbation theory below $\sim
30\,\gev$. The continuum extrapolation of the results for three of the
definitions of the RGI mass are shown in
\fig{fig:mc}, along with the results of other calculations. The
discretization errors displayed by the VWI and $\bar cc$-AWI results
are of the expected size $\sim (am_c)^2$, and the RGI masses obtained
from the three definitions extrapolate linearly in $a^2$ to surprisingly
consistent values in the continuum limit, at $a=0$. Comparable agreement
is found with the two other definitions. In the continuum limit the
authors of \cite{Rolf:2002gu} quote
$\mmsbar_c(\mmsbar_c)=1.301(28)(20)(7)\,\gev$,
using the $N^3LL$ results of \cite{Chetyrkin:1997dh,Vermaseren:1997fq} 
to convert the
RGI mass to the $\msbar$ scheme. The first error is a combination of
statistical and a number of systematic uncertainties; the second is due
to the uncertainty in $\Lambda^{(0)}_\msbar$ \cite{Capitani:1998mq};
the third is the difference of the results obtained using $N^2LL$ and
$N^3LL$ expressions for the conversion from the RGI mass. In addition,
a quenched scale ambiguity of $10\%$ is found to induce a 3\%
uncertainty on $\mmsbar_c(\mmsbar_c)$.

Before concluding on the charm-quark mass, let me mention that Juge
{\it et al.} \cite{Juge:2001dj} are performing a modern version of
Kronfeld's preliminary 1997 calculation \cite{Kronfeld:1997zc}, where
the spin-averaged mass of the $1S$ $c\bar c$-states was used to tune
$m_c$ and the lattice spacing was determined from the $1P-1S$
splitting. In this new calculation, performed at several lattice
spacings, the mass is renormalized perturbatively at $N^2LO$ instead
of at $NLO$.

\smallskip

My conclusion for the charm-quark mass is that the result of
\cite{Rolf:2002gu} {\it is} the quenched result. So I use it as
the central value and add to it a 15\% systematic error to account for
quenching uncertainties. Thus, I quote as a lattice summary number,

\smallskip
\ovalbox{
\begin{minipage}{6.7cm}
\vspace{-0.1cm}
$$
\mmsbar_c(\mmsbar_c)=1.30(4)(20)\,\gev
\ .
$$
\end{minipage}
}

\smallskip
\noindent
Because of the large quenching uncertainty, this result does not
improve on the 2002 PDG summary, $1.0\,\gev\le \mmsbar_c(\mmsbar_c)\le
1.4\,\gev$, which does not include lattice results
\cite{Hagiwara:pw}. It is clear that the error on $m_c$ from the lattice can
only be reduced now with an {\it unquenched} calculation.

\subsection{The $b$-quark mass}

Since the $b$-quark mass is the $B$-meson mass up to a smallish
correction, what we really want is an accurate calculation of this
correction.  This statement is all the more relevant that the $b$-quark
mass cannot be reached in large volumes on present day lattices. Thus
we rely on the heavy-quark expansion of the $B$-meson mass:
\be
\begin{array}{rccl}
M_B & = & \mbare_b  &+ \cE + \ord{\frac{\lqcd^2}{m_b}}\\
& = & \mpole_b-\delta m &+ \cE  +\ord{\frac{\lqcd^2}{m_b}}
\end{array}
\ ,
\label{eq:hqexpofmb}
\ee
where the residual mass $\delta m$ and the bare binding energy $\cE$ are
linearly divergent individually, but this divergence cancels in the 
difference.

What the heavy-quark expansion seeks to achieve is a separation of
short and long-distance contributions to $M_B$. In \eq{eq:hqexpofmb},
$\cE$ is meant to contain the long-distance components of $M_B$ and
$m_b^{bare}$, the short-distance contributions.  This division can be
made unambiguous with a hard cutoff, such as the lattice spacing, and
$\cE$ can be computed in the static approximation on lattices which
are not particularly fine.

Two approaches to the determination of $m_b$ have been pursued in the
context of HQET: a
``perturbative'' one \cite{Crisafulli:1995pg} and a
non-perturbative one. The latter was proposed in the last year
\cite{Heitger:2001ch,Sommer:2002en} and represents an important
theoretical development in the matching of HQET to QCD.

\subsubsection{The ``perturbative'' approach}

In this approach, the determination of $m_b$ proceeds in three steps:

\smallskip

a) $\cE$ is evaluated numerically in the static approximation;

\smallskip

b) $\delta m$ is calculated in lattice-HQET perturbation theory; it is
known to three loops for $N_f=0$ 
\cite{DiRenzo:2000nd,Trottier:2001vj} and two loops for $N_f\ne 0$
\cite{Martinelli:1998vt};

\smallskip

c) the $b$-quark mass is then obtained through
$$
\mmsbar_b(\mmsbar_b)=c(\mmsbar_b)(M_B+\delta m-\cE)+
\mbox{\small$\ord{\frac{\lqcd^2}{m_b}}$}
\ ,
$$
where $c(\mmsbar_b)$ is known to three-loop accuracy
\cite{Melnikov:2000qh}.

\smallskip

The problems with this approach are:

\smallskip

$\bullet$ that the cancellation of power divergences between $\cE$ and
$\delta m$ is not complete: $\cE -\delta m \sim \alpha_s^n/a$ ($n{=}3$
for N$^2$LO calculations) which becomes large in the continuum limit;

\smallskip

$\bullet$ $\delta m$ has a renormalon ambiguity of order $\lqcd$ which
cancels against the one in $c(\mmsbar_b)$
\cite{Beneke:1994sw,Bigi:1994em}.

\smallskip

Both these problems require one to go to high order in perturbation
theory. That this is not only a conceptual problem can be inferred
from the fact that the uncertainty which the neglect of higher-order
terms induces on $\mmsbar_b(\mmsbar_b)$ is about 200~MeV at NLO,
100~MeV at N$^2$LO and 50~MeV at N$^3$LO, for $N_f=0$.

The first problem further means that a continuum limit is not possible.
Thus, one is left with discretization errors proportional to powers
of $a\lqcd$ and perturbative errors of order $\alpha_s^3$ 
for N$^2$LO calculations.

\subsubsection{The non-perturbative approach}

This approach was proposed very recently by Heitger and Sommer
\cite{Heitger:2001ch,Sommer:2002en}. 
It provides {\it a framework in which to match HQET and QCD
fully non-perturbatively and in the continuum limit.} In that sense it
solves all of the problems of the perturbative method.

Since this development is interesting conceptually, it is worth trying
to get the main idea across. It is, however, rather technical and I
will only provide a sketch here:

\smallskip

a) First observe that the bare quark mass~\footnote{The light-quark-mass
dependence of $\cE$ and $M$ is suppressed for clarity.}
\be
\mbare(\mrgi,a) =  M(L_0,\mrgi,a)-\cE(L_0,a)
\label{eq:mbare}
\ee
$$+\ord{1/L_0^2\mrgi,\lqcd/L_0\mrgi,\lqcd^2/\mrgi}
\ ,
$$
which is a parameter in the HQET lagrangian, knows nothing about 
$L_0$, the size
of the lattice considered. It is only a function of the RGI quark mass
and the lattice spacing $a$. In \eq{eq:mbare}, 
$M(L_0,\mrgi,a)$ is a quantity with mass-dimension one which becomes
the mass $M$ of the heavy-light meson containing a heavy quark
of mass $\mrgi$ in the limit $L_0\to\infty$ and $a\to 0$. 

The independence of $\mbare$ on $L_0$ means that when $\mrgi\sim
\mrgi_b$, it can be studied on small $(1/\lqcd\ge L_0\gg 1/\mrgi_b)$
and fine-grained $(a \mrgi_b\ll 1)$ lattices where discretization
errors proportional to powers of $\sim a\mrgi_b$ are small.

\smallskip

b) The second step consists in equating $\mbare(\mrgi,a)$, defined through
\eq{eq:mbare}, on a small ($L_0$) and large ($L$) box:
\be
M\simeq \cE(L,a)-\cE(L_0,a)+M(L_0,\mrgi,a)
\ ,
\label{eq:mbarelL0L}
\ee
where I have assumed that $L$ is large enough and $a$ small enough
for $M(L,\mrgi,a)$ to be the meson mass $M$ to good approximation. It
is this mass which is going to be set to the experimental value
of the $B$-meson mass to determine $\mrgi_b$.

\smallskip

c) Zero is then inserted
into the right-hand side of \eq{eq:mbarelL0L}, in the following way:
\bea
M&\simeq& \cE(L,a) -
\cE(L_n,a)+\cdots-\cE(L_1,a) \label{eq:mbarelL0Ln}\\
&+&\cE(L_1,a)-\cE(L_0,a)+M(L_0,\mrgi,a)\nn
\ ,
\eea
where one takes $L=L_{n+1}> L_n=2^n L_0 >\cdots > L_0$. The trick is to note
that one can consider a sequence of lattice spacings, $a\le
a_0 \le\cdots\le a_n\ll 1/\lqcd$, and that as long as $a_i/L_i\ll 1$,
\eq{eq:mbarelL0Ln} is equivalent to
\bea
M&\simeq& \hspace{-0.3cm}\cE(L,a_n) -
\cE(L_n,a_n)+\cdots-\cE(L_1,a_1) \label{eq:mbarelL0Lnan}\\
&+&\cE(L_1,a_0)-\cE(L_0,a_0)+M(L_0,\mrgi,a)\nn
\ ,
\eea
up to small discretization errors proportional to powers of $a_i/L_i$
and $a_i\lqcd$. This breaks up the problem of having to accommodate
long-distance physics on a scale $L$ and short-distance effects
associated with the scale $1/\mrgi_b$ into a sequence of more manageable
problems. Indeed, for each binding energy difference
$\Delta\cE_i=\cE(L_{i+1},a_i)-\cE(L_i,a_i)$, $i=0,\cdots, n$, 
the power divergences
$\sim 1/a_i$ in $\cE(L_{i+1},a_i)$ and $\cE(L_i,a_i)$ cancel and the
range of scales involved, from $a_i$ to $L_{i+1}$, is not too large.

\begin{figure}
\centerline{\epsfxsize=4.cm\epsffile{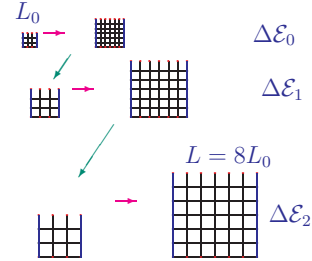}}
\vspace{-1.0cm}
\caption{\em Sequence of lattices on which the binding-energy differences, 
$\Delta\cE_i$, are computed.}
\vspace{-0.5cm}
\label{fig:recursion}
\end{figure}

\smallskip

d) This means that the $\Delta\cE_i$ can be computed at leading order
in lattice HQET on the sequence of lattices illustrated in
\fig{fig:recursion}. Furthermore, in each step of the sequence, 
the continuum limit
$a_i\to 0$ can be taken.

\smallskip

e) Next compute $M(L_0,\mrgi,a)$ in lattice QCD for $\mrgi\sim\mrgi_b$,
taking $a\to 0$. This is possible, since the range of scales involved
$a\ll 1/\mrgi_b\ll L_0$ is not too large. Then, interpolate $M(L_0,\mrgi,0)$
to the value of $\mrgi$ which solves
\be
M(L_0,\mrgi,0) = M_{B}-\sum_{i=0}^n \Delta\cE_i
\ .\ee
This yields $\mrgi=\mrgi_b+\ord{1/L_0^2\mrgi_b,\cdots}$. $\mrgi_b$ can
be then converted to $\mmsbar_b$ using $N^3LL$ running 
\cite{Chetyrkin:1997dh,Vermaseren:1997fq} 
if necessary. The
authors of \cite{Heitger:2001ch} choose $L_0\simeq 0.2\,\fm$, $n=2$,
$L\simeq 1.5\fm
\sim 2^3 L_0$ and use $M_{B_s}$ instead of $M_B$ to tune to $\mrgi_b$.

\smallskip

The great benefit of this approach is that $\mrgi_b$ is obtained
without discretization nor perturbative errors, in contrast to the
``perturbative'' approach described earlier. Power corrections to the
heavy-quark limit, however, are still present (see
\eq{eq:mbare}). Those involving $1/L_0\mrgi_b$ are peculiar to this
finite-volume approach and their size should be investigated
carefully.

\smallskip

Other approaches, based on lattice NRQCD and a study of upsilon
\cite{Davies:1994pz,Hornbostel:1998ki} or heavy-light \cite{AliKhan:1999yb} 
spectra, have also been explored. These calculations were performed
using NLO perturbation theory and are therefore less accurate than
present day calculations. 

\smallskip

\begin{figure}
\centerline{\epsfxsize=6cm\epsffile{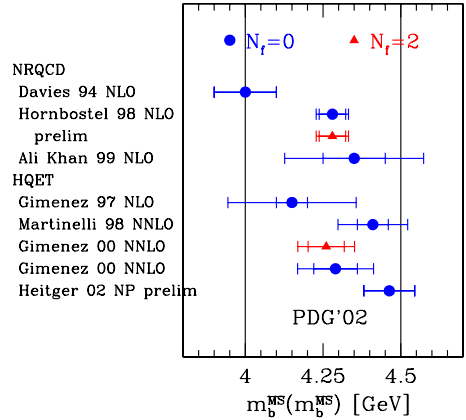}}
\vspace{-1.cm}
\caption{\em 
Lattice results for the $b$-quark mass: 
Davies~94 \cite{Davies:1994pz}, Hornbostel~98
\cite{Hornbostel:1998ki}, Ali Khan~99 \cite{AliKhan:1999yb}, 
Gimenez~97 \cite{Gimenez:1996av},
Martinelli~98 \cite{Martinelli:1998vt}, Gimenez~00
\cite{Gimenez:2000cj}, Heitger~02
\cite{Heitger:2001ch,Sommer:2002en}. Next to the reference is the
order in perturbation theory used in the calculation. NP stands for
non-perturbative.  The vertical lines delimit the range given by the
PDG \cite{Hagiwara:pw}.}
\vspace{-0.7cm}
\label{fig:Mb_summary}
\end{figure}

Having presented the various approaches used to determine the
$b$-quark mass on the lattice, I now review the results
obtained. These results are summarized in \fig{fig:Mb_summary}. Only
published results are presented. The exception being the only fully
non-perturbative result of \cite{Heitger:2001ch,Sommer:2002en} and the
unquenched, $N_f=2$, result of \cite{Hornbostel:1998ki}, the latter
because it enables an estimate of quenching uncertainties.

Not shown are preliminary quenched, N$^3$LO results which were
obtained using NRQCD and an extrapolation to the infinite-mass limit
\cite{sara,Collins:2000sb}. At this same N$^3$LO order, the result of
a preliminary quenched HQET calculation was reported in
\cite{Lubicz:2000ch}. Both these results are compatible with the
N$^2$LO results of
\cite{Gimenez:2000cj}, but have a perturbative
uncertainty about twice as small ($\sim 50\,\mev$).

The only unquenched results available \cite{Gimenez:2000cj,
Hornbostel:1998ki,Collins:2000sb} show
little change compared to the corresponding quenched results.
However, the masses of the sea quarks considered are rather large.  In
\cite{Gimenez:2000cj}, for instance, the calculation is performed 
for two values of the sea-quark mass, $m_{sea}$, in the interval
$m_s/2$ to $m_s$. These findings should be checked through further
unquenched studies.

Performing a simple average of the quenched $\beta=6.2$ result of
\cite{Gimenez:2000cj} and the preliminary non-perturbative result of
\cite{Heitger:2001ch,Sommer:2002en} yields

\smallskip
\ovalbox{
\begin{minipage}{6.7cm}
\vspace{-0.1cm}
 $$\mmsbar_b(\mmsbar_b)=4.38(9)(10)\,\gev \ ,$$
\end{minipage}
}

\smallskip
\noindent
where the first error is typical of the quenched calculations
considered, chosen also to accommodate the two values within one standard
deviation, and the second is an estimate of the remaining quenching
uncertainty, taken to be 10\% of the $B_s$ binding energy. The
quenching error was reduced from the 15\% used for the charm mass to
10\% because of the evidence provided by the $N_f=2$ results discussed
above.

\section{Lattice QCD for the unitarity triangle}

\label{sec:ckmology}

Analyses of the CKM unitarity triangle (UT) have come to rely more and
more on weak matrix elements calculated in lattice QCD (please see
Stocchi's plenary review \cite{achille} and e.g. \cite{fabrizio,
Hocker:2001xe}). In these analyses, the summit $(\bar\rho,\bar\eta)$
of the UT is constrained through $\Delta m_d$, the frequency of
$B^0_d-\bar B^0_d$ oscillations, $\Delta m_d/\Delta m_s$, the ratio of
this frequency to the one for $B^0_s-\bar B^0_s$ oscillations,
$\epsilon_K$ which parametrizes indirect $CP$ violation in
$K\to\pi\pi$ decays and the ratio $|V_{ub}/V_{cb}|$, together with the
value of $\sin 2\beta$ determined from the time-dependent $CP$
asymmetry in $B\to J/\psi K_S$. The theoretical expressions for
$\Delta m_d$, $\Delta m_s/\Delta m_d$ and $\epsilon_K$ involve CKM
factors and short-distance Wilson coefficients which are known to NLO
in perturbation theory (see e.g. \cite{Buchalla:1995vs} for a
review). They also contain the non-perturbative QCD quantities
$f_{B_d}^2 B_{B_d}$, $\xi^2=\frac{f_{B_s}^2 B_{B_s}}{f_{B_d}^2
B_{B_d}}$ and $B_K$. These quantities are defined through the matrix
elements:
\bea
\la\bar B_q|(\bar bq)_{V-A}(\bar bq)_{V-A}|B_q\ra
=\frac83 M_{B_q}^2f_{B_q}^2 B_{B_q}
\ ,&&
\label{eq:db2me}
\\
\la 0|\bar b\gamma_\mu\gamma_5q|B_q(p)\ra=ip_\mu f_{B_q}
\ ,\hspace{2.35cm}&&
\label{eq:fbme}
\\
\la\bar K^0|(\bar sd)_{V-A}(\bar sd)_{V-A}|K^0\ra
=\frac{8}{3} M_K^2 f_K^2 B_K
\ ,&&
\label{eq:ds2me}
\eea
with $q=d$, $s$. It is for the calculation of these matrix elements
that lattice QCD is used. 

In addition to the matrix elements mentioned above, lattice QCD can
also be used to obtain the form factors relevant for $B\to
D^*(D)\ell\nu$ and $B\to
\pi(\rho)\ell\nu$. Together with these form factors, experimental
measurements of the corresponding differential decay rates enable
independent determinations of $|V_{cb}|$ and $|V_{ub}|$, respectively.

\subsection{$B_q-\bar B_q$--mixing: $f_{B_q}$}

Though $f_{B_q}$ and $B_{B_q}$ are correlated through the $\Delta B=2$
matrix element of \eq{eq:db2me}, it is useful to review these
quantities separately: once measured, leptonic decays of $B^+$, which
are governed by $f_B$, will allow a clean determination of $|V_{ub}|$;
there exist many more calculations of $f_{B_q}$; systematics on
$f_{B_q}$ and $B_{B_q}$ are quite different.

\begin{figure}[t]
\centerline{\epsfysize=5.5cm\epsffile{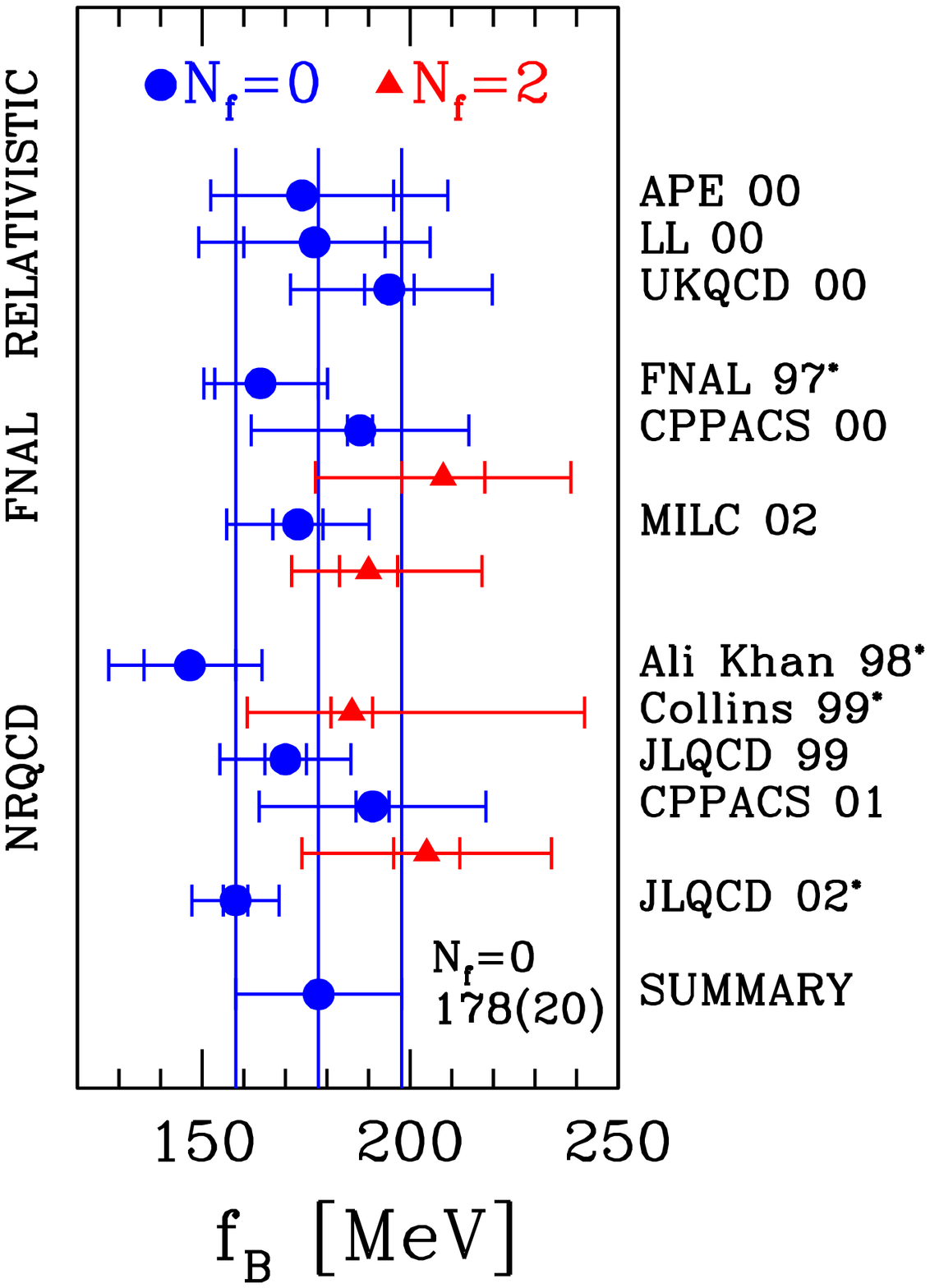} 
\epsfysize=5.5cm\epsffile{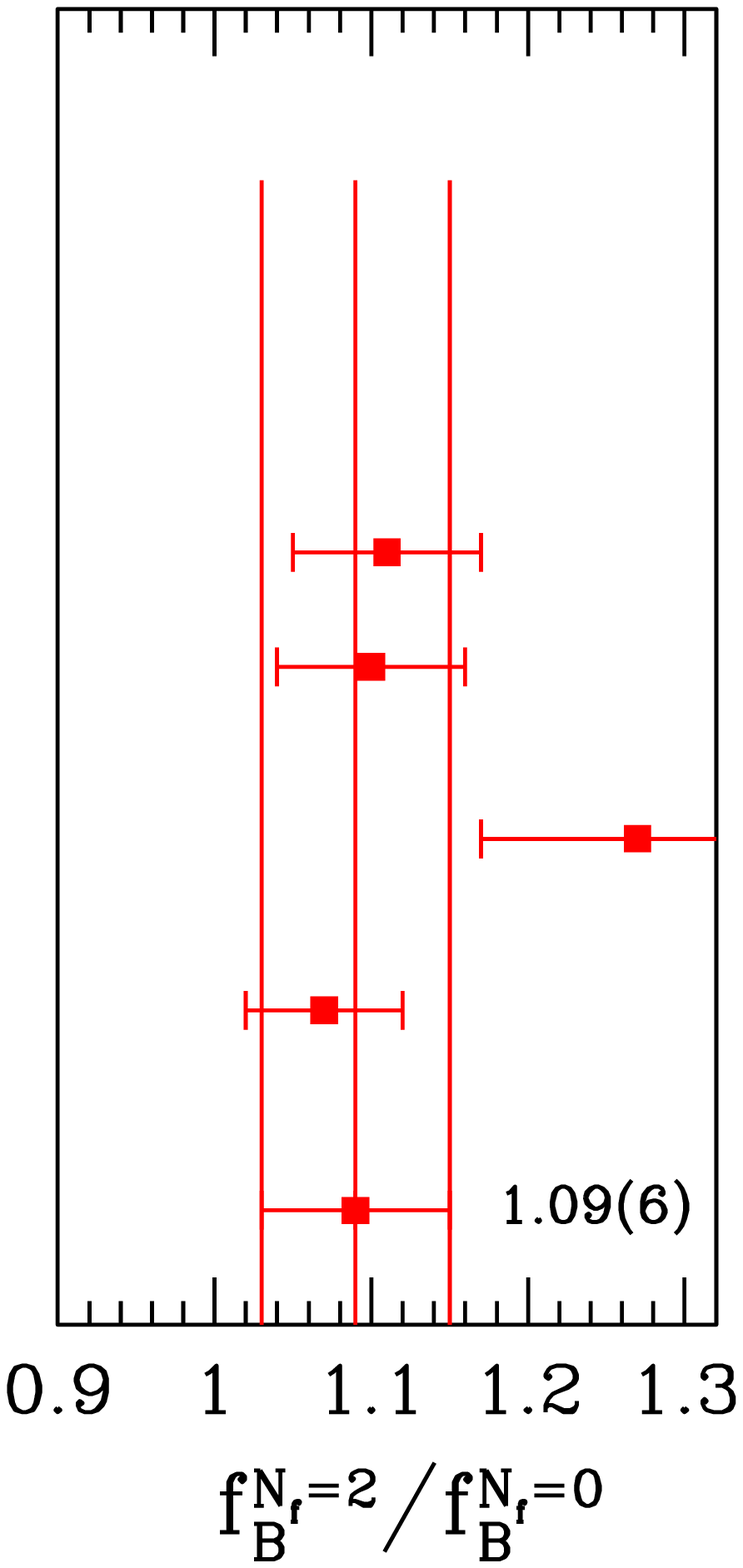}}
\vspace{-1.cm}
\caption{\em Lattice results for the decay constant of the $B$ meson
in quenched ($N_f=0$) and unquenched QCD ($N_f=2$) (left) and their
ratio (right). Calculations are grouped by the approach used for
the heavy quark. Only results published after 1996 are shown. These
are: APE~00 \cite{Becirevic:2000nv}, LL~00 \cite{Lellouch:2000tw},
UKQCD~00 \cite{Bowler:2000xw}, FNAL~97
\cite{El-Khadra:1997hq}, CP-PACS~00 \cite{AliKhan:2000eg}, MILC~02
\cite{Bernard:2002pc}~\protect\footnotemark, Ali Khan~98
\cite{AliKhan:1998df}, Collins~99
\cite{Collins:1999ff}, CP-PACS~01 \cite{AliKhan:2001jg}, 
JLQCD~02 \cite{Yamada:2002wh}. Also shown are my averages of the
results for $f_B^{N_f=0}$ and $f_B^{N_f=2}/f_B^{N_f=0}$. 
}
\vspace{-0.5cm}
\label{fig:fB}
\end{figure}

\begin{figure}[t]
\centerline{\epsfysize=5.5cm\epsffile{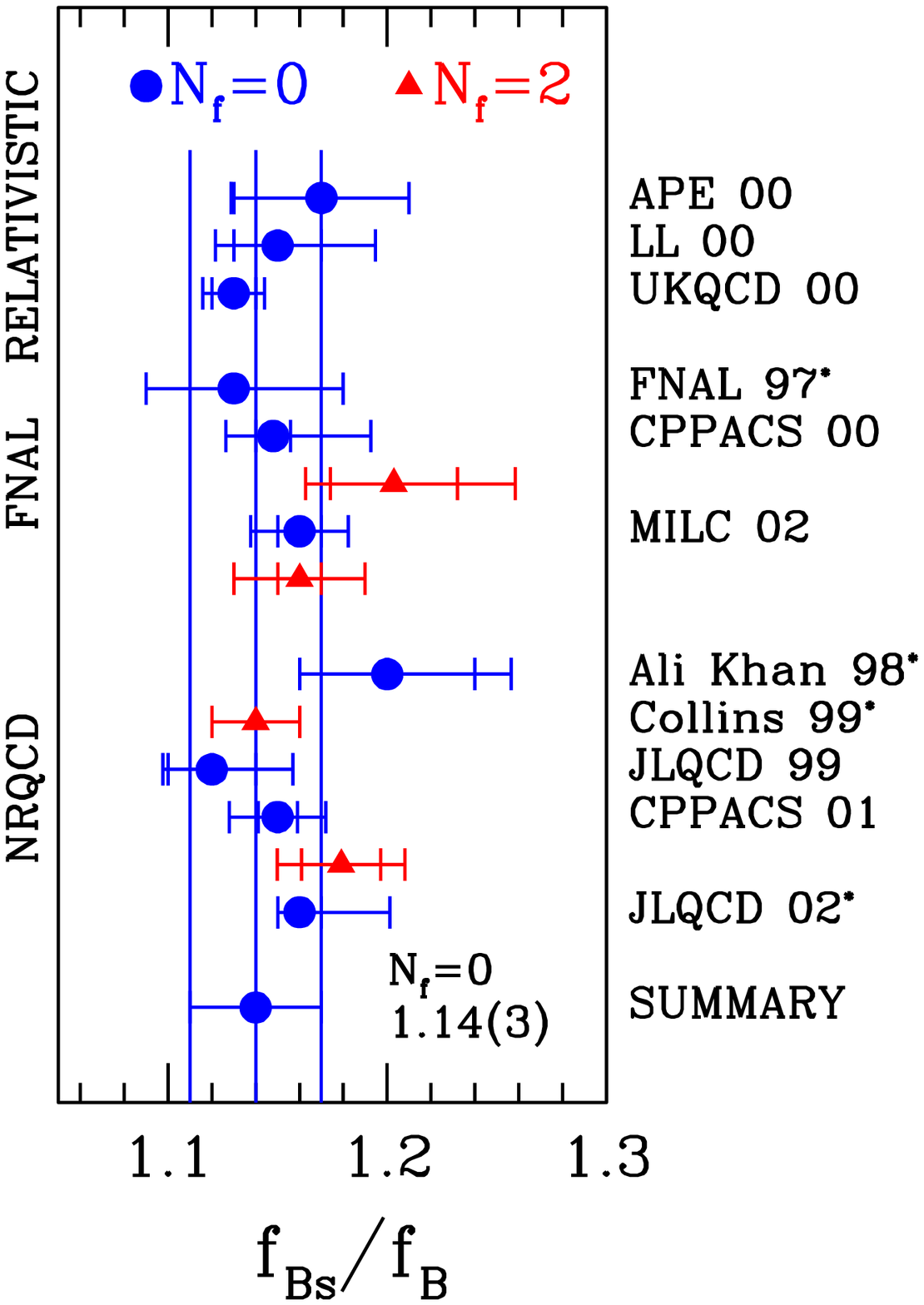} 
\epsfysize=5.5cm\epsffile{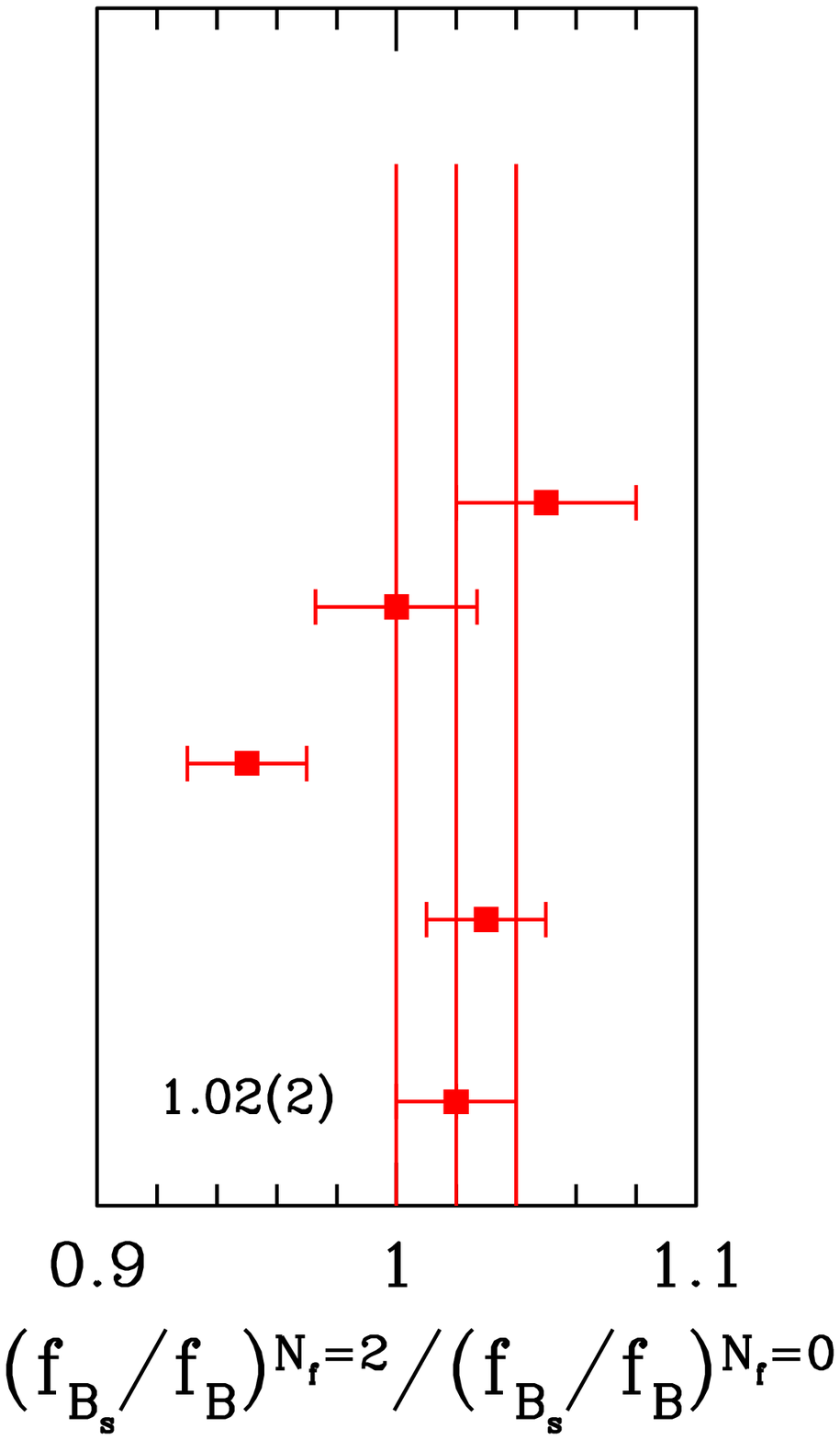}}
\vspace{-1.cm}
\caption{\em As in \fig{fig:fB}, but for $f_{B_s}/f_B$ and
$(f_{B_s}/f_B)^{N_f=2}/(f_{B_s}/f_B)^{N_f=0}$.  
}
\vspace{-0.5cm}
\label{fig:fBsfB}
\end{figure}

In \fig{fig:fB}, I have compiled all recent results for
$f_B$~\footnote{Here and below, $f_B\equiv f_{B_u}=f_{B_d}$ 
and $B_B\equiv B_{B_d}$.}, both quenched and
unquenched. A similar compilation for $f_{B_s}/f_B$ is shown in
\fig{fig:fBsfB}.
\footnotetext{The results of \cite{Bernard:2002pc} 
include an additional systematic error to account for the
possible effect of chiral logarithms. I have not included it in 
\figs{fig:fB}{fig:fBsfB} because none of the other results presented
consider these effects 
and because I will discuss them separately in \sec{sec:fbxext}.}
From these figures it is clear that results obtained with different
approaches to treating heavy quarks (relativistic QCD {\em vs} effective
theories) agree fairly well, suggesting that the heavy-quark-mass
dependence of these quantities is under control. 

The figures also indicate that $N_f=2$ calculations show an increase
in $f_B$ over its quenched value of up to $\ord{15\%}$ while these
effects appear to be small in $f_{B_s}/f_B$. These trends are
confirmed by MILC who recently presented the first
results of a calculation performed with three flavors of sea quarks
($N_f=3$) \cite{Bernard:2001wy,Bernard:2002ep}. Their preliminary
findings are $f_B^{N_f=3}>f_B^{N_f=2}$ and $f_B^{N_f=3}/f_B^{N_f=0}=
1.23(4)(6)$ and $(f_{B_s}/f_B)^{N_f=3}=1.18(1)^{+4}_{-1}$.

Before embarking on a discussion of systematic errors, I wish to
briefly mention that an interesting finite-volume technique for
computing $B$-meson decay constants was presented very recently
\cite{Guagnelli:2002jd}.  The first quenched results 
obtained using this method are $f_B=170(11)(5)(22)\,\mev$ and
$f_{B_s}=192(9)(5)(24)\,\mev$ \cite{Guagnelli:2002jd}.

\subsubsection{$f_{B_q}$: chiral extrapolations}

\label{sec:fbxext}

In the quenched approximation, the calculation of $f_B$ requires a
chiral extrapolation in light-valence-quark mass from $m_{val}\gsim
m_s/2$ to $m_{u,d}$.  In unquenched calculations, an additional chiral
extrapolation in sea-quark mass from $m_{sea}\gsim m_s/2$ to $m_{u,d}$
is needed to obtain both $f_B$ and $f_{B_s}$.  All of the results
presented in the figures above assume mild extrapolations. It is
legitimate to ask whether the chiral behavior of $f_B$ and $f_{B_s}$
is well understood. This is a serious issue for UT fits, due to the
importance of the $\Delta m_s/\Delta m_d$ constraint.

According to $N_f=2$ heavy meson chiral perturbation theory
(HM$\chi$PT), the light-quark-mass behavior of $f_B$, where the mass
of light valence and sea quarks are identified ($m_{val}=m_{sea}$), is
given by \cite{Grinstein:1992qt}
{\small
\be
\frac{f_B\sqrt{M_B}}{\phi_B^{(0)}}=
1-\frac34(1+3g^2)\l[\frac{M_{SS}}{4\pi f}\r]^2
\ln\frac{M_{SS}^2}{\Lambda_{f_B}^2},
\label{eq:fbxpt}
\ee
}
\noindent
at leading order in the $1/M_B$ expansion and up to NNLO chiral
corrections. In \eq{eq:fbxpt}, $\phi_B^{(0)}$ is the value of
$f_B\sqrt{M_B}$ in the combined heavy-quark and chiral limits, $g$ is
the $B^*B\pi$ coupling in this same limit, $M_{SS}$ is the mass of a
pseudoscalar meson composed of a quark and anti-quark of mass
$m_{sea}$, $f$ is the pion decay constant in the chiral limit and
$\Lambda_{f_B}$ is a scale which parametrizes the unknown coupling of
the contact term proportional to $M_{SS}^2$.~\footnote{At the order
in which
\eq{eq:fbxpt} is obtained, $g$ cannot be distinguished from the physical
$B^*B\pi$ coupling, $g_b$, and $f$ from $f_\pi$ or even $f_{SS}$.} For
$f_{B_s}$ and $N_f=2$, one has to resort to partially quenched
HM$\chi$PT, since the valence and sea light quarks are not
degenerate. The prediction for the sea-quark-mass dependence of
$f_{B_s}$ is \cite{Sharpe:1995qp} ($m_{val}$ fixed to $m_s$)
{\small
\be
\frac{f_{B_s}\sqrt{M_{B_s}}}{\phi_{B_s}^{(0)}}=
1-(1+3g^2)\l[\frac{M_{VS}}{4\pi f}\r]^2
\ln\frac{M_{VS}^2}{\Lambda_{f_{B_s}}^2},
\label{eq:fbsxpt}\ee
}
\noindent
at leading order in the $1/M_B$ expansion and up to NNLO
chiral corrections. In \eq{eq:fbsxpt}, $M_{VS}$ is the mass of a
pseudoscalar meson composed of quarks of mass $m_{val}=m_s$ and
$m_{sea}$. The dependence on sea-quark mass is recovered through the
LO $\chi$PT relation, $M_{VS}^2=(M_{SS}^2+M_{VV}^2)/2$.  This
dependence is much milder for $f_{B_s}$ than it is for $f_B$, because
$M_{VS}$ is much less sensitive to $m_{sea}$ than is $M_{SS}$.

CLEO has recently determined $g_c=0.59(7)$ using its measurement of
the $D^{*+}$ width. $g_c$ is the $D^*D\pi$ coupling and differs from
$g$ by chiral and $1/M_D$ corrections. This value is consistent with a
recent lattice calculation which further finds little dependence of
the $D^*D\pi$ coupling on the charm quark mass
\cite{Abada:2002vj}. However, agreement amongst different
determinations of $g$ is generally rather poor, the spread of which
lies in the interval $0.2<g<0.7$
\cite{Colangelo:2002dg,Colangelo:2000dp,Becirevic:1999fr}. 

\begin{figure}[t]
\centerline{\epsfxsize=7.cm\epsffile{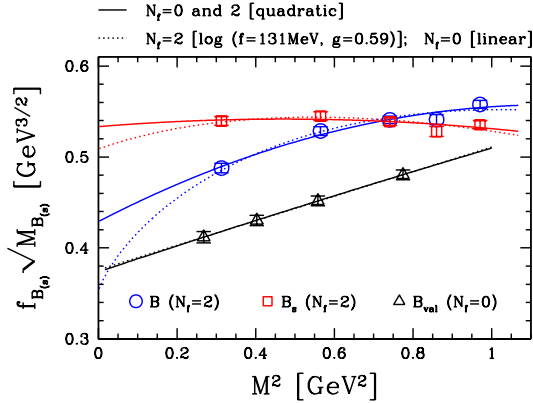 }}
\vspace{-1.cm}
\caption{\em Preliminary results for the dependence of
$f_{B_{(s)}}\sqrt{M_{B_{(s)}}}$ ($f_{B_{val}}\sqrt{M_{B_{val}}}$) on
light-quark (valence-quark) mass in $N_f=2$ ($N_f=0$) NRQCD
\cite{Hashimoto:2002vi} (\cite{nori_ckm}). $M\equiv M_{SS}$ for
$N_f=2$ and $M\equiv M_{VV}$ when $N_f=0$. The curves denoted log
correspond to fits to
\eqs{eq:fbxpt}{eq:fbsxpt}.  
}
\vspace{-0.7cm}
\label{fig:fbfbsxextrap}
\end{figure}

JLQCD have achieved an impressive statistical accuracy for their $N_f=2$
determinations of $f_B$ and $f_{B_s}$ \cite{Hashimoto:2002vi}. This
has enabled them to study rather extensively the dependence of these
quantities on light-quark mass.~\footnote{A direct comparison of 
the JLQCD results to those of MILC
\cite{Bernard:2002pc}, which have larger statistical errors, is
difficult: the two sets of results are obtained at fixed
bare coupling, with different quark actions, in different volumes, with
different procedures for setting the scale, in addition to which the
masses of the $b$ and $s$ quarks in the JLQCD calculation are not yet
precisely tuned to their physical values. However, a rough
comparison indicates that the JLQCD and MILC results display similar
features in the region of light-quark mass where they overlap.} 
These preliminary studies are
illustrated in
\fig{fig:fbfbsxextrap}, where the 
quantities $f_{B}\sqrt{M_B}$ and $f_{B_s}\sqrt{M_{B_s}}$ are plotted
against $M_{SS}^2$. In the quenched case, where no sea quarks are
present, $f_{B_{val}}\sqrt{M_{B_{val}}}$ is a function of
$M_{VV}^2$. In fact, the near linearity of these quenched results,
observed in all such calculations, is one of the reasons why
chiral extrapolations have been done assuming mild
light-quark-mass dependence.

JLQCD perform a number of fits to their results for $f_B\sqrt{M_B}$
and $f_{B_{s}}\sqrt{M_{B_{s}}}$ versus $M_{SS}^2$, including fits to
the HQ$\chi$PT expressions of \eqs{eq:fbxpt}{eq:fbsxpt}
\cite{Hashimoto:2002vi}.~\footnote{What is actually plotted in 
\fig{fig:fbfbsxextrap}, for $N_f{=}2$, is
$r_0^{3/2}f_{B_{(s)}}\sqrt{M_{B_{(s)}}}$ {\em vs} $(r_0M_{SS})^2$,
with the replacement $r_0\to 2.47\,\gev^{-1}$, as obtained by
rescaling $r_0$ by a long-distance observable close to the chiral
limit. The physical units are therefore only meant to be indicative.
The assumption behind fitting $f_{B_{(s)}}\sqrt{M_{B_{(s)}}}$, thus
obtained, to the HQ$\chi$PT expressions of
\eqs{eq:fbxpt}{eq:fbsxpt} is that $r_0$ does not 
have a sea-quark-mass dependence
which could significantly modify the coefficient of the logs in these
equations. This is not unreasonable since the scale $r_0$ is defined
through the force between static quarks at $r_0$ and is found to be
$\sim 0.5\,\fm$ \cite{Sommer:1993ce}. At such distances, this force
should not be very sensitive to pions.} For $f_{B_s}\sqrt{M_{B_s}}$,
\fig{fig:fbfbsxextrap} indicates that the extrapolated value at
$M_{SS}^2=M_\pi^2$ is affected only mildly by the presence of chiral
logs. The situation is radically different for
$f_{B}\sqrt{M_{B}}$ where the extrapolated value depends critically on
the coupling $g$, becoming lower for larger values of $g$.  For
$g=0.6$, JLQCD find that the extrapolated value of $f_{B}\sqrt{M_{B}}$
is 17\% lower than that obtained through a quadratic fit in $M_{SS}^2$
\cite{Yamada:2002wh}.  They conclude that the inclusion of logarithms
in these chiral extrapolations could significantly lower the value of
$f_B$ and increase that of $f_{B_s}/f_B$.  Similar conclusions were
reached about $f_{B_s}/f_B$ by the authors of
\cite{Kronfeld:2002ab}, using the predicted $N_f=3$ chiral logs and 
information from lattice calculations. Based on these observations
Yamada, who reviewed heavy-quark physics at Lattice 2002, chose to
quote for $f_B$ a central value taken from linear/quadratic chiral
extrapolations, adding to it a $-17\%$ systematic error to account for
possible effects of chiral logs
\cite{Yamada:2002wh}. Given the size of this error and its effect on
UT fits, I wish to make a number of comments on the JLQCD analyses:

\smallskip

{$\bullet$} The chiral log extrapolation of the results of JLQCD for
$f_B\sqrt{M_{B}}$ is problematic from a numerical analysis 
point of view in the
sense that this functional form varies most rapidly in a 
region unconstrained by data.

\smallskip

{$\bullet$} The size of the NLO chiral correction of \eq{eq:fbxpt} is
$\ord{60\%}$ at the heaviest point. With such a sizeable correction,
higher-order terms cannot be ignored. Moreover,
$\chi$PT is not expected to hold up to $1\,\gev$ since the expansion is
in $M_{SS}^2$ over $8\pi^2 f^2\simeq (1.1\,\gev)^2$.

\smallskip

{$\bullet$} Another important piece of information is the behavior of
$f_\pi$ {\em vs} $M_\pi^2$ obtained by JLQCD with 
the same gauge configurations
\cite{Hashimoto:2002vi}. While
the coefficient of the chiral log in the $\chi$PT expression for
$f_\pi$ \cite{Gasser:1983yg} may even be larger than the one for
$f_B\sqrt{M_B}$ (2 {\em vs} the 1.6 obtained for $g=0.6$), their
nearly linear results for $f_\pi$ are inconsistent with NLO chiral
behavior ($\chi^2/dof=5.1$) \cite{Hashimoto:2002vi} in the range of
sea quark masses where they claim that logarithmic behavior may have
been observed for $f_B\sqrt{M_B}$.

\smallskip

{$\bullet$} Because JLQCD work at fixed bare gauge coupling and with a
fixed number of lattice sites, the spacing and the volume of their
lattice shrink by roughly 25\% as $M_{SS}^2$ varies from $\sim
1\,\gev^2$ to $\sim 0.3\,\gev^2$. This rather large variation will
amplify the unphysical dependence on $M_{SS}^2$ induced by
mass-dependent finite-volume and discretization errors.

\smallskip

{$\bullet$} If instead of requiring that the NLO behavior of
\eq{eq:fbxpt} be applicable all the way up to $M_{SS}^2\sim
1\,\gev^2$, one models the possible effect of higher-order chiral
terms, allowing or not the logarithmic behavior to set in at lower
masses, one finds that the downward shift in JLQCD's results for $f_B$
due to chiral logs is generically less than about 10\% (for
$g\le 0.6$). For $f_{B_s}$, a similar analysis gives variations which
are negligible compared to other systematics.

\smallskip

It is important to note, also, that the uncertainty due to chiral
logs depends on the details of the calculation and, in
particular, on whether the lattice spacing was set with a quantity
which is sensitive to chiral logs and whether these logarithms were
taken into account.  This means that one cannot generically ascribe a
chiral-log error to quantities such as $f_B$. However, the quenched
results reviewed above, which form the basis of the summary numbers
presented below, include an error associated with scale setting and
agree within errors despite the fact that different groups set the
scale with quantities which have varying degrees of sensitivity to
chiral logs. This indicates that the chiral-log uncertainty due to
scale setting is to some extent already accounted for. The same
appears to be true for the $N_f=2$ over $N_f=0$ ratios which are also
used to obtain the final summary numbers.  Thus, I will assume that
the discussion surrounding the chiral behavior of the JLQCD results
can be used to obtain a rough estimate of the range of deviations that
could be induced by the chiral logs intrinsic to $f_B$ and
$f_{B_s}$.  Given that discussion, the exploratory nature of the
investigations of chiral-log effects, it seems reasonable to fix the
central values of $f_B$ and $f_{B_s}$ from linear/quadratic chiral
extrapolations and to add a $-10\%$ systematic error to $f_B$ to
account for the uncertainty in the chiral extrapolation and none to
$f_{B_s}$.

To reduce the chiral-log error in future analyses, the obvious
solution is to extend the lattice results to smaller masses to match
unambiguously onto NLO chiral behavior. This is numerically very
costly, however.~\footnote{Nonetheless, MILC are making impressive
progress in that direction \cite{Bernard:2002ep}.} In the meantime, it
would be helpful to chirally extrapolate dimensionless ratios of
quantities in which the chiral logs cancel or nearly cancel. For
such ratios, the uncertainties due to contributions of chiral
logarithms at small light-quark mass will be much reduced. For $f_B$
in $N_f{=}2$ calculations, and $g$ around 0.6, a good
candidate would be $f_B/f_\pi$ (see discussion above and also
\cite{Bernard:2002pc}).~\footnote{At leading order in the $1/M_B$
expansion, $f_B$ and $f_B\sqrt{M_B}$ have identical chiral
expansions.} To make such analyses reliable, it is important to treat
chiral logarithms consistently throughout the calculation
\cite{Bernard:2002pc}, not to consider light quarks which are too
massive, to reduce the uncertainty on $g$ and to determine the extent
to which $1/M_B$ corrections can modify the prediction of
\eq{eq:fbxpt}. It is also important to verify that variations
in lattice spacing and volume such as those mentioned above do not
distort dependences on light-quark mass significantly.

\subsection{$B_q-\bar B_q$--mixing: $B_{B_q}$}

\begin{figure}[t]
\centerline{\epsfxsize=6.cm\epsffile{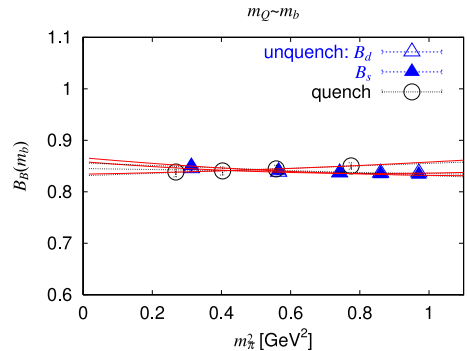}}
\vspace{-1.cm}
\caption{\em Dependence of
$B_B$ and $B_{B_s}$ on light-quark (valence-quark) mass 
in $N_f=2$ ($N_f=0$)
NRQCD \cite{nori_ckm}. The light-quark mass
is parametrized by the mass squared of the pseudoscalar meson composed of 
quarks with that quark mass. The different curves correspond
to linear (dotted) and quadratic (solid) fits.}
\vspace{-0.6cm}
\label{fig:bbxextrap}
\end{figure}

There are many fewer calculations of $B_{B_q}$ than there are of $f_{B_q}$.
However, the situation with chiral extrapolations and with unquenching
appears to be much more favorable. Indeed, for $N_f=2$, the
coefficient of the chiral logarithm for $B_B$ is in the
range $-0.2<\frac12(1-3g^2)<0.4$ instead of $0.8<\frac34(1+3g^2)<1.9$ as
it is for $f_B$, when $0.2<g<0.7$. For $g=0.6$, these
coefficients are $0.0$ and $1.6$, respectively. This milder behavior
is confirmed by the unquenched calculation of JLQCD 
\cite{Yamada:2001xp,nori_ckm}, whose results for the light-quark-mass
dependence of $B_B$ and $B_{B_s}$ are shown in
\fig{fig:bbxextrap}.~\footnote{The fact that the chiral behavior of
the decay constants may be affected by the variation of lattice
spacing and volume mentioned above does not imply that the
$B$-parameters are (see discussion around
\eq{eq:bbdef}).} This mild light-quark-mass dependence is also
observed in all quenched calculations
\cite{Lellouch:2000tw,Becirevic:2000nv,
Bernard:1998dg,Hashimoto:2000eh,Aoki:2002bh} of which an example is
shown in
\fig{fig:bbxextrap}. Also evident from \fig{fig:bbxextrap} is the fact
the there is very little variation in going from $N_f=0$ to $N_f=2$,
indicating that quenching effects are small for $B_B$ and $B_{B_s}$.

The situation is less clear when it comes to the heavy-quark-mass
behavior of $B_{B_q}$. In \cite{Becirevic:2000nv,Lellouch:2000tw}, $B$
parameters computed in the relativistic approach for heavy-quark
masses around that of the charm are extrapolated up to the $b$, using
HQET as a guide. Leading logs are accounted for in
\cite{Lellouch:2000tw}. In
\cite{Becirevic:2001xt}, NLL corrections in the matching of QCD to
HQET are included, and a calculation of $B_{B_q}$ in the static
limit is used in conjunction with the relativistic results of
\cite{Becirevic:2000nv} to reach the $b$-quark by interpolation.
These three calculations indicate that $B_{B_q}$ increases with
increasing $M_{B_q}$. On the other hand, the NRQCD results of
\cite{Aoki:2002bh}, performed at quark masses above and around $m_b$, 
suggest an opposite behavior. This behavior is
mild, however, and the different formulations agree at the physical
$b$-quark mass. Moreover, once systematics errors are included, the
significance of these inconsistencies becomes marginal.

\begin{figure}[t]
\centerline{\epsfysize=5.cm\epsffile{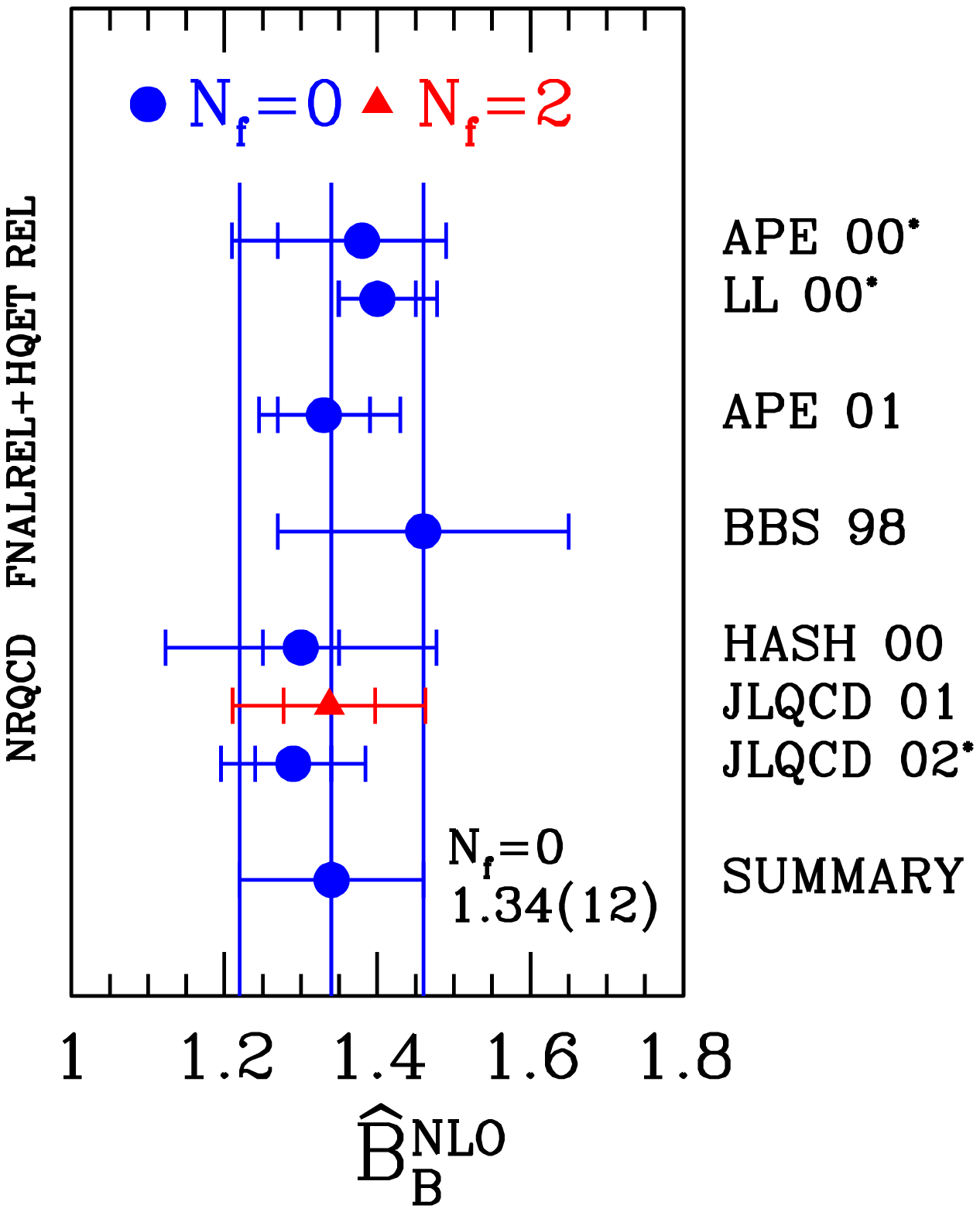}
\epsfysize=5.cm\epsffile{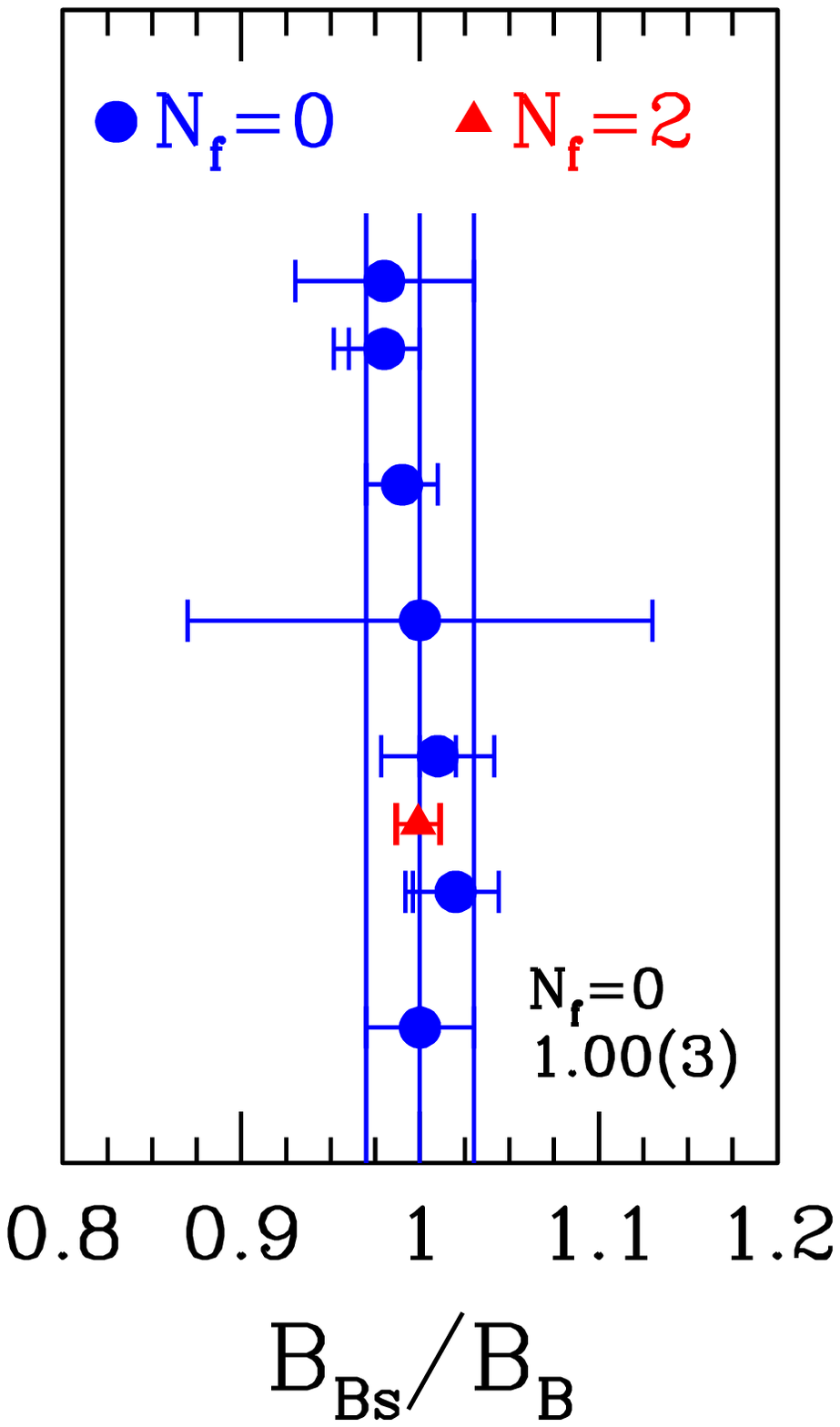}}
\vspace{-1.cm}
\caption{\em Lattice results for $\bnlo_B$ and $B_{B_s}/B_B$. 
Calculations are grouped by the approach used for
the heavy quark. The results shown are: APE~00 \cite{Becirevic:2000nv}, 
LL~00 \cite{Lellouch:2000tw},
APE~01 \cite{Becirevic:2001xt}, BBS~98
\cite{Bernard:1998dg}, HASH~00 \cite{Hashimoto:2000eh}, 
JLQCD~01 \cite{Yamada:2001xp}, JLQCD~02 \cite{Aoki:2002bh}.
Also shown are my averages of the quenched results for these quantities.
}
\vspace{-0.6cm}
\label{fig:bbandbbsres}
\end{figure}

The results of calculations of the NLO-RGI parameter, $\bnlo_B$, are
collected in \fig{fig:bbandbbsres}, together with those for $B_{B_s}/B_B$.
All methods give fully consistent results suggesting that residual
systematic errors cancel in the ratio of matrix elements, 
\be
B_{B_q}=\frac38\frac{\la\bar B_q|(\bar bq)_{V-A}(\bar bq)_{V-A}|B_q\ra}
{\la \bar B_q|\bar q\gamma^\mu\gamma_5b|0\ra
\la 0|\bar b\gamma_\mu\gamma_5q|B_q\ra}
\ ,
\label{eq:bbdef}
\ee
used to define the $B$-parameters. This appears also to be true of
quenching effects.  

The mild dependence on light valence-quark mass further guarantees that
errors on $B_{B_s}/B_B$ are small, of order 3\%. The situation regarding
the heavy-quark-mass dependence, however, should be clarified. This could
be achieved by combining relativistic and HQET results obtained with
non-perturbative renormalization in the continuum
limit (see \sec{sec:hqanddisc}).

\subsection{$B_q-\bar B_q$--mixing: summary}

I provide here summary numbers for the non-perturbative quantities
relevant for the study of $B_q-\bar B_q$--mixing. In performing
averages, I have only kept results dated after 1998 in the
quenched case ($N_f=0$) and after 1999 for the unquenched case
($N_f=2$) and omit results which had not made it into proceedings or
papers at the time of the conference. I use quenched averages as a
starting point and the ratio of $N_f=2$ to $N_f=0$ results to
extrapolate these results to $N_f=3$, including an additional
systematic error equal to the shift from $N_f=2$ to $N_f=3$ in the
final result. I obtain

\smallskip
\ovalbox{
\begin{minipage}{6.7cm}
\vspace{-0.3cm}
$$
f_B = 203(27)^{+0}_{-20}\,\mev,\ f_{B_s}=238(31)\,\mev,
$$
$$
\bnlo_B = 1.34(12),\quad \bnlo_{B_s}=1.34(12),
$$
$$
f_B\sqrt{\bnlo_B} = 235(33)^{+0}_{-24}\,\mev,
$$
$$
f_{B_s}\sqrt{\bnlo_{B_s}}=276(38)\,\mev,
$$
\end{minipage}
}

\smallskip
\noindent
and
\smallskip

\ovalbox{
\begin{minipage}{6.7cm}
\vspace{-0.3cm}
$$
\frac{f_{B_s}}{f_B}=1.18(4)^{+12}_{-0},\quad
\frac{B_{B_s}}{B_B}=1.00(3),
$$
$$
\xi\equiv \frac{f_{B_s}\sqrt{B_{B_s}}}{f_B\sqrt{B_B}}=
1.18(4)^{+12}_{-0}, 
$$
\end{minipage}
}
\smallskip

\noindent
where the second, asymmetric error is the one due to the uncertainty
in the chiral extrapolation of $f_B$ discussed above. This error
should be viewed as rough estimate of the range of possible deviations
that can be induced by including chiral logs in this
extrapolation. Thus, in UT analyses, the value of $\xi$ given above
should be understood as $\xi=1.24(4)(6)$~\footnote{This result is
entirely consistent with the recent phenomenological analysis of
\cite{Becirevic:2002mh}.} and similarly for the other quantities
affected by this error. The results were not presented in this way
because the inclusion of chiral logs is still exploratory. Note that
this discussion does not concern $f_{B_s}$ and
$f_{B_s}\sqrt{\bnlo_{B_s}}$ whose residual chiral-log uncertainties
appear to be small compared to other errors.

Before closing, let me mention that two groups have calculated $\xi$
directly from the ratio of $\Delta B=2$ matrix elements
\cite{Bernard:1998dg,Lellouch:2000tw}. 
The results obtained come out higher than the 1.18 given above, but
are compatible within errors.

\subsection{$K^0-\bar K^0$--mixing and $B_K$}

The standard model, $\Delta S=2$ matrix element relevant for $CP$
violation in $K^0-\bar K^0$--mixing has been studied extensively
on the lattice. A feature common to all but one of the calculations
reviewed below is the use of the quenched approximation.~\footnote{The
exception is the preliminary $N_f=3$ result of
\cite{Kilcup:1997hp}.} Furthermore, in these calculations, kaons are
composed of mass-degenerate quarks and the final result is obtained by
interpolation to a kaon of mass $M_K$, made up of two quarks of mass $\sim
m_s/2$. The procedure has the advantage that a chiral extrapolation to
the mass of the down quark is avoided. The drawback is that
$SU(3)$-breaking corrections $\sim (m_s-m_d)^2$ are not included and
must be accounted for in the systematic error.

\subsubsection{New calculations of $B_K$}

This year three new quenched calculations of $B_K$ have been
performed. The first, with Wilson type fermions, was
presented at this conference
\cite{damir,Becirevic:2002mm}. It is performed with high statistics 
at three different values of the lattice spacing in the range from
$0.1$ to $0.06$ fermi, allowing for a continuum extrapolation. It
implements a new method based on chiral Ward identities to eliminate
the spurious mixing with wrong chirality four-quark operators
\cite{Becirevic:2000cy} as well as the more traditional method of
chiral subtractions. The authors find good agreement between their two
preliminary results in the continuum limit. They renormalize the
$B$-parameter non-perturbatively in the RI/MOM scheme.

The two other studies \cite{gghlr02,DeGrand:2002xe} were performed
using overlap quarks \cite{Neuberger:1997bg}, a recent discretization
of QCD which satisfies the Ginsparg-Wilson relation \cite{Ginsparg:1981bj} 
and thus exhibits
an exact chiral-flavor symmetry, analogous to the one of continuum
QCD, at finite lattice spacing \cite{Luscher:1998pq}. This symmetry
implies that the $\Delta S=2$ operator is simple to construct on the
lattice (unlike with staggered fermions) and does not mix with
chirally dominant operators
\cite{Hasenfratz:1998jp} (as it does with
Wilson fermions). Overlap fermions further guarantee full
$\ord{a}$-improvement.  These calculations are the first weak matrix
element studies to use overlap quarks. They help validate this
discretization as a useful phenomenological tool despite its present
numerical cost. Such calculations will eventually verify that the
explicit breaking of flavor (staggered fermions) or chiral (Wilson
fermions) symmetry is controlled in standard calculations. They will
also provide a check of domain wall fermion calculations. Domain wall
quarks \cite{Kaplan:1992bt,Shamir:1993zy} are a five-dimensional
implementation of Ginsparg-Wilson fermions. The chiral-flavor symmetry
which they exhibit becomes exact in the limit of infinite fifth
dimension.

The first of these two overlap fermion calculations \cite{gghlr02} is
performed in the quenched approximation at a single value of the
lattice spacing ($a^{-1}\sim 2\,\gev$) for five values of the
light-quark mass in the range from $m_s/2$ to $1.3 m_s$. The
$B$-parameter is renormalized non-perturbatively in the RI/MOM scheme.

The second overlap calculation \cite{DeGrand:2002xe} is also performed
in the quenched approximation, at two values of the lattice spacing
($a^{-1}\sim 1.7\,\gev$ and $2.2\,\gev$), with a slightly different
action and with five light-quark masses, ranging from $0.8 m_s$ to
$2.5m_s$. In the mass interval common to both calculations, agreement
is excellent. However, the large quark masses considered in
\cite{DeGrand:2002xe} imply that the physical point is
reached through a somewhat dangerous (linear) extrapolation. The
resulting $B$-parameter is renormalized at one-loop.  Results at
the two lattice spacings are in good agreement.

Before summarizing the situation regarding $B_K$, I wish to mention
that ALPHA \cite{Guagnelli:2002rw} are performing a
high-statistics calculation of this $B$-parameter using 
non-perturbative renormalization and a modified version of Wilson
fermions which goes under the name of twisted-mass QCD
\cite{Frezzotti:2000nk}. This approach circumvents the problem of
mixing with wrong chirality operators in much the same way as the Ward
identity method of
\cite{Becirevic:2000cy}, which it inspired.

\subsubsection{Summary of $B_K$ determinations}

\begin{figure}[t]
\centerline{\epsfxsize=7.cm\epsffile{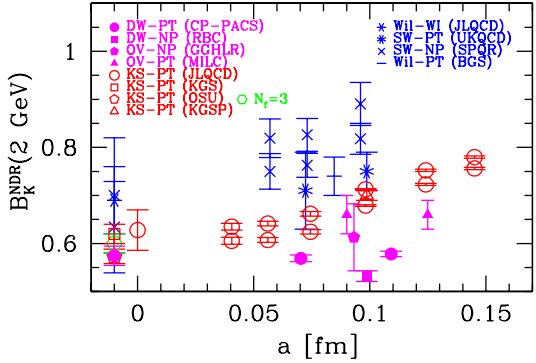}}
\vspace{-1.cm}
\caption{\em Lattice 
results for $\bndr_K(2\,\gev)$ {\em vs} lattice spacing. The results
are coded according to the quark action used: Ginsparg-Wilson (filled
symbols), staggered or Kogut-Susskind (unfilled symbols) and Wilson
(stick symbols). The continuum limit corresponds to $a=0$. All but the
reference result of JLQCD \cite{Aoki:1997nr} in that limit are shifted
to the left for clarity.  The results shown are: CP-PACS
\cite{AliKhan:2001wr}, RBC \cite{Blum:2001xb}, GGHLR \cite{gghlr02},
MILC \cite{DeGrand:2002xe}, JLQCD (KS)
\cite{Aoki:1997nr}, KGS \cite{Kilcup:1997ye}, OSU and $N_f=3$ 
\cite{Kilcup:1997hp}, 
KSGP \cite{Kilcup:1990fq}, SPQR \cite{Becirevic:2002mm}, 
JLQCD (Wilson) \cite{Aoki:1999gw}, UKQCD \cite{Lellouch:1999sg}, 
BGS \cite{Gupta:1997yt}. Not shown are the pioneering domain wall
results of \cite{Blum:1997mz}.}
\vspace{-0.7cm}
\label{fig:bksummary}
\end{figure}

Results for $\bndr_K(2\,\gev)$ obtained by various groups using
different approaches are plotted as a function of lattice spacing in
\fig{fig:bksummary}. While results differ at finite lattice spacing,
those which are extrapolated to the continuum limit agree in this
limit within one-and-some standard deviations. The overlap results of
\cite{gghlr02,DeGrand:2002xe}, obtained at finite lattice spacing, are also 
in good agreement with these continuum results, albeit with large
statistical errors. Domain wall results are a bit on the low
side. This difference should be understood through further scrutiny
of systematic errors in those calculations.

The reference result is still the one obtained in the 1997 quenched
staggered calculation of JLQCD
\cite{Aoki:1997nr}, which is the most complete study of this quantity
to date, despite its use of one-loop perturbation theory in matching
the lattice operators to their continuum counterparts.  To their
result one must add uncertainties due to quenching and the fact that
the down and strange quarks are degenerate in the calculation. Sharpe
\cite{Sharpe:1998hh}, on the basis of quenched $\chi$PT and the
preliminary ``$N_f=3$'' OSU results
\cite{Kilcup:1997hp}, suggests an enhancement factor of $1.05\pm 0.15$
to ``unquench'' quenched results for $B_K$. He advocates an additional
factor $1.05\pm 0.05$ to compensate for the effect of working with
degenerate quark masses. Here, I choose to keep his estimate of
errors, but not to use his enhancement factors so as not to modify the
central value on the basis of information which is provided by $\chi$PT
estimates and preliminary lattice results. Thus, I quote
%


\ovalbox{
\begin{minipage}{6.7cm}
$$
\bndr_K(2\,\gev)=0.628(42)(99)$$
$$
\to \bnlo_K=0.86(6)(14)\ ,
$$
\end{minipage}
}
%

\noindent
where $\bnlo_K$ is the two-loop RGI $B$-parameter obtained from
$\bndr_K(2\,\gev)$ with $N_f=3$ and $\alpha_s(2\,\gev)=0.3$. 

This is the same result as was presented at Lattice 2000
\cite{Lellouch:2000bm}. What is really needed now are high statistics
unquenched calculations to reduce the dominant 15\% quenching
error. Without such an improvement in accuracy, indirect $CP$ violation
in the neutral kaon system will no longer provide a significant
constraint on UT fits \cite{fabrizio}.

\subsection{Semileptonic decays}

\subsubsection{$|V_{cb}|$ from $B\to D^*\ell\nu$}

The CKM element $|V_{cb}|$ plays an important rôle in constraining the
UT, for it determines the triangle's base but also appears raised to
the fourth power in the constraint on the summit provided by
$\epsilon_K$. $|V_{cb}|$ must thus
be determined as precisely as possible.

One way to measure $|V_{cb}|$ accurately is by extrapolating
the differential decay rate for $B\to D^*\ell\nu$ decays,
$$
\frac{d\Gamma}{dw}\sim |V_{cb}|^2\,|\cF_{D^*}(w)|^2\ ,
$$
to the zero-recoil point $w=v_B\cdot v_{D^*}=1$, where $v_B$ and
$v_{D^*}$ are the four-velocities of the initial and final mesons
\cite{Neubert:td}. The benefit of such an extrapolation is that heavy quark
symmetry \cite{Shifman:1987rj,Isgur:vq} requires that
the form factor $\cF_{D^*}(w)$ must be normalized to 1 at $w=1$, up to
small radiative and small order $1/m_{b,c}^2$ corrections
\cite{Luke:1990eg}. Nevertheless, a
few percent measurement of $|V_{cb}|$ requires a reliable
determination of $\cF_{D^*}(1)-1$. While the radiative corrections
have been calculated to two-loops in perturbation theory
\cite{Czarnecki:1998wy}, the long-distance power corrections require a
non-perturbative treatment.

Through a clever use of double ratios of matrix elements for
$D^{(*)},B^{(*)}\to D^{(*)},B^{(*)}$--like transitions, the authors of
\cite{Hashimoto:2001nb} have been able to obtain a statistically
significant signal for the leading non-perturbative corrections of
order $1/m_{b,c}^2$ and some of the sub-leading ones proportional to
three factors of the inverse heavy-quark masses. The result of their
quenched calculation, performed at three values of the lattice
spacing, was presented at this conference
\cite{jim,Hashimoto:2001nb}:
$$
\cF_{D^*}(1)=0.913^{+24+17}_{-17-30}
\ .
$$
This important result, which requires excellent control of statistical
and systematic errors, should be verified by other lattice groups.

\subsection{$B\to\pi\ell\nu$ and $|V_{ub}|$}

In heavy-to-light semileptonic decays, the light, final state hadron
can have momenta as large as $|\vec{p}|\sim m_Q/2$ in the parent rest
frame, where $m_Q$ is the mass of the heavy quark. For $m_Q=m_b$, and
on present day lattices, such momenta would lead to uncontrollably
large discretization effects, proportional to powers of
$a|\vec{p}|$. Therefore, at present, only a limited kinematical range
below the zero-recoil point can be reached without extrapolation. Even
so, lattice calculations are useful, for the relevant matrix elements
are not normalized by heavy quark symmetry as they are in $B\to
D^{(*)}\ell\nu$ decays.  Furthermore, experiment is beginning to
measure the corresponding differential rates in the kinematical region
accessible to lattice calculations (see e.g. \cite{kwon}), thus allowing
model-independent determinations of $|V_{ub}|$.

\begin{figure}[t]
\centerline{\epsfxsize=7.2cm\epsffile{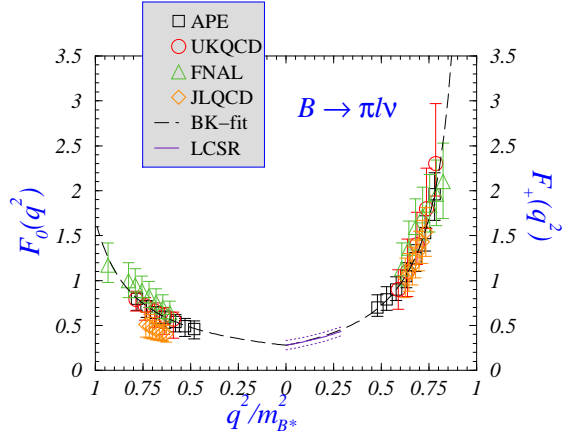}}
\vspace{-1.2cm}
\caption{\em Lattice results for 
$B^0\to\pi^-\ell^+\nu$ form factors $F_+(q^2)$ and $F_0(q^2)$ (figure
from \cite{damir}). These results are APE \cite{Abada:2000ty}, 
UKQCD \cite{Bowler:1999xn}, FNAL
\cite{El-Khadra:2001rv}, JLQCD \cite{Aoki:2001rd}. 
Also shown is a fit to the parametrization of
\cite{Becirevic:1999kt} and the light-cone sumrule results
of \cite{Khodjamirian:2000ds} (see also \cite{Ball:2001fp}).}
\vspace{-0.5cm}
\label{fig:bpisummary}
\end{figure}

The matrix element relevant for $B^0\to\pi^-\ell^+\nu$ decays is
\bea
&\la\pi^-(p')|V^\mu|B^0(p)\ra = \frac{M_B^2-M_\pi^2}{q^2}q^\mu F_0(q^2)&\nn\\
&+(p+p'-\frac{M_B^2-M_\pi^2}{q^2}q)^\mu F_+(q^2)\ ,&\nn
\eea
where $q=p-p'$ and $V^\mu=\bar b\gamma^\mu u$. There are four quenched
calculations of this matrix element, using either relativistic, Fermilab
or NRQCD heavy quarks. The form factors $F_+(q^2)$ and $F_0(q^2)$
obtained by the different collaborations are plotted as functions
of $q^2$ in \fig{fig:bpisummary}. Excellent agreement is found for
$F_+(q^2)$ which determines the rate in the limit of vanishing lepton
mass. This agreement is certainly not trivial given the use of
different heavy-quark approaches and the variations in the rather
intricate analyses leading to the determination of these form
factors. The error on this form factor is typically of order 20\%.

The lattice results cover 
roughly the upper half of the kinematically accessible range. The
easiest way to extrapolate these results to lower values of $q^2$ is
to use a parametrization such as the one presented in
\cite{Becirevic:1999kt}, which incorporates most of the known
constraints on the form factors. Such an extrapolation is shown in
\fig{fig:bpisummary} and gives excellent agreement with light-cone
sumrule results \cite{Khodjamirian:2000ds,Ball:2001fp}. This
extrapolation introduces, however, a model dependence. To eliminate
this model dependence, one can make use of the dispersive bound techniques
proposed in \cite{Boyd:1994tt,Lellouch:1995yv}. In fact, the results of the
analysis of \cite{Lellouch:1995yv} could certainly be improved with
the lattice results presented here.

Given the good agreement amongst the different quenched calculations
of $F_+(q^2)$, what is needed to reduce the error on the extraction of
$|V_{ub}|$ are unquenched calculations and more work on methods for
enlarging the kinematical reach of lattice calculations.

\section{Outlook}

\label{sec:ccl}

A number of quantities of central importance to particle physics are
currently being studied with lattice QCD, many of which could
unfortunately not be presented here. For those that were, emphasis is,
in most cases, on the reduction of systematic errors, such as
quenching, discretization errors, etc. 

Because they have not yet led to phenomenological breakthroughs, I did
not discuss in any detail the major advances of the last few years
associated with the formulation and implementation of fermionic
discretizations which preserve an exact continuum-like chiral-flavor
symmetry at finite lattice spacing.  These discretizations are known
as Ginsparg-Wilson fermions \cite{Ginsparg:1981bj}, of which overlap
\cite{Neuberger:1997bg}, domain wall \cite{Kaplan:1992bt} and
fixed-point fermions \cite{Hasenfratz:1997ft} are explicit
realizations. They have opened new possibilities for the calculation
of weak matrix elements, and in particular those associated with the
$\Delta I=1/2$ rule and direct $CP$ violation in $K\to\pi\pi$ decays. On
the analytical side, the renormalization of the weak operators
involved has been studied with domain wall \cite{Aoki:2000ee} and
overlap fermions
\cite{Capitani:2000bm,Capitani:2000da}. On the numerical side, a first
round of investigations has been performed with domain wall fermions
\cite{Noaki:2001un,Blum:2001xb}, laying the foundations 
for future studies.

The exceptional chiral properties of these new fermion formulations
has also allowed the investigation of an unexplored regime of QCD in
which the correlation length of the pion field is much larger than the
linear extent of spacetime, the $\epsilon$-regime of
\cite{Gasser:1987ah}. This investigation has lead to 
exploratory calculations of one of the low-energy constants
(LEC) of the strong chiral lagrangian in the quenched approximation
\cite{Hernandez:1999cu,Hernandez:2001yn,
DeGrand:2001ie,Hasenfratz:2002rp}. It is possible, in principle, to
generalize this approach to extract the LECs of the weak chiral
lagrangian and a numerical investigation is under way \cite{kpipi}.

More generally, the range of approaches and of quantities studied in
lattice QCD is constantly expanding and there should be many more
exciting results to present at ICHEP 2004.

\bigskip

\noindent
{\bf Acknowledgments}

\medskip

I thank D.~Becirevic, C.~Bernard, J.~Charles, S.~Collins, J.~Flynn,
L.~Giusti, S.~Hashimoto, V.~Lubicz, G.~Martinelli, F.~Parodi, S.~Sint,
R.~Sommer, A.~Stocchi and N.~Yamada for useful discussions and for
their help in preparing this review. I would also like to thank the
LOC, and Eric Laenen in particular, for their logistic support and
more generally for making ICHEP 2002 a very interesting and 
enjoyable conference.

\input{proc.bbl}
\end{document}

%% file: macros.tex
\newcommand{\lsim}{ {\
\lower-1.2pt\vbox{\hbox{\rlap{$<$}\lower5pt\vbox{\hbox{$\sim$}}}}\ } }
\newcommand{\gsim}{ {\
\lower-1.2pt\vbox{\hbox{\rlap{$>$}\lower5pt\vbox{\hbox{$\sim$}}}}\ } }

\def\fm{\mathrm{fm}}
\def\ev{\mathrm{e\kern-0.1em V}}
\def\kev{\mathrm{ke\kern-0.1em V}}
\def\mev{\mathrm{Me\kern-0.1em V}}
\def\gev{\mathrm{Ge\kern-0.1em V}}
\def\tev{\mathrm{Te\kern-0.1em V}}
\def\ra{\rangle}

\def\la{\langle}
\def\l{\left}
\def\r{\right}
\def\ord#1{\mathcal{O}\l(#1\r)}
\def\be{\begin{equation}}
\def\ee{\end{equation}}
\def\bea{\begin{eqnarray}}
\def\eea{\end{eqnarray}}
\def\nn{\nonumber}

\def\cF{\mathcal{F}}

\def\cE{\mathcal{E}}
\def\lqcd{\Lambda_\mathrm{QCD}}
\def\msbar{{\overline{\mbox{MS}}}}
\def\msbartiny{{\overline{\mbox{\tiny MS}}}}

\def\rgitiny{{\mbox{\tiny RGI}}}

\def\mrgi{m^{\rgitiny}}
\def\mbare{m^{\mbox{\tiny bare}}}
\def\mpole{m^{\mbox{\tiny pole}}}
\def\mmsbar{m^\msbartiny}
\def\bnlo{\hat B^{\mbox{\tiny NLO}}}
\def\bndr{B^{\mbox{\tiny NDR}}}

\def\eq#1{Eq.\ (\ref{#1})}
\def\eqs#1#2{Eqs.\ (\ref{#1}) and (\ref{#2})}

\def\sec#1{Sec.\ \ref{#1}}

\def\fig#1{Fig.\ \ref{#1}}
\def\figs#1#2{Figs.\ \ref{#1} and \ref{#2}}


%% file: titlehep.tex
\title{
{
\vspace{-3.0cm} \normalsize \hfill
\parbox{30mm}{ CPT-2002/P.4443} 
}\\[22mm]
\vspace{-0.7cm}
Phenomenology from lattice QCD\thanks{Plenary talk at {\em ICHEP 2002},
24-31 July 2002, Amsterdam, The Netherlands. Work supported in 
part by EC HPP contract HPRN-CT-2000-00145 (Hadrons/Lattice QCD) 
and HPRN-CT-2002-00311 (EURIDICE).\newline
$^\ddag$ Unité Propre de Recherche 7061.
}}

\author{Laurent Lellouch\address{Centre de Physique Théorique~$^\ddag$, 
Case 907, CNRS Luminy, F-13288 Marseille Cedex, France}%
}